\documentclass{article}
\usepackage{graphicx}
\usepackage{amsmath}
\usepackage{amsfonts}
\usepackage{amssymb}

\begin{document}
\begin{titlepage}
\begin{flushright}
CAMS/02-02\\
\end{flushright}
\vspace{.3cm}
\begin{center}
\renewcommand{\thefootnote}{\fnsymbol{footnote}}
{\Large\bf An Invariant action for Noncommutative Gravity in Four-Dimensions}%

\vskip20mm {\large
\bf{A.
H. Chamseddine \footnote{email: chams@aub.edu.lb} }}\\
\renewcommand{\thefootnote}{\arabic{footnote}}
\vskip2cm { Center for Advanced Mathematical Sciences (CAMS) \\and\\ Physics
Department\\American University of Beirut\\Beirut,  Lebanon\\}
\end{center}
\vfill
\begin{center}
{\bf Abstract}
\end{center}
\vskip.5cm
Two main problems face the construction of noncommutative actions for
gravity with star products: the complex metric and finding an invariant measure.
The only gauge groups that could be used with star products are the unitary
groups. I propose an invariant gravitational action in $D=4$ dimensions based
on the constrained gauge group $U(2,2)$  broken to
$U(1,1)\times U(1,1).$ No metric is used, thus giving a naturally
invariant measure. This action is generalized to
the noncommutative case by replacing ordinary products with star
products. The four dimensional noncommutative action is studied and
the deformed action to first order in deformation parameter is computed.
\end{titlepage}

In noncommutative field theory based on the Moyal star product \cite{CDS},
\cite{SW} the only gauge theories that can be used are based on unitary
algebras. The presence of a constant background B-field for open or closed
strings with D-branes lead to the noncommutativity of space-time coordinates.
The Einstein-Hilbert action can be constructed either by insuring
diffeomorphism invariance or local Lorentz invariance \cite{Utiyama}%
,\cite{Kibble}. This program faces difficulties when ordinary products are
replaced with star products. In this case, it is not an easy matter to define
a generalization of Riemannian geometry . Noncommutative Riemannian geometry
has been developed for noncommutative spaces based on the spectral triple
\cite{CFF},\cite{Connes}. The difficult part in applying this formalism is to
determine the deformed spectral triple. In particular, the deformed Dirac
operator is needed in order to apply this formalism to noncommutative spaces
where the algebra is deformed with the star product. One must also find an
invariant measure. There is, however, some recent progress on such formulation
\cite{CD}. Recently, the effective action for gravity on noncommutative branes
in presence of constant background B-field was derived and found to be
non-covariant \cite{SL}. This conforms to the expectation that in this case
space-time coordinates do not commute.

The approach based on gauging the Lorentz algebra also have problems, mainly
that the metric becomes complex, and the antisymmetric part of the metric may
have non-physical propagating modes \cite{complex}. Finding an invariant
measure is also problematic in this approach. \ One way to avoid the problem
of finding an invariant measure is to require the action to be an invariant
D-form in a D-dimensional space \cite{cho},\cite{CW}. Experience with building
gauge invariant actions which are also D-forms in a D-dimensional space tells
us that these actions are usually topological, and therefore cannot describe
gravity in dimensions of four or higher \cite{topology}. This is usually
avoided by imposing constraints on some components of the gauge field
strengths which, in some cases, is equivalent to a torsion free metric theory
\cite{van}. Constraints insure that the action, although metric independent,
is not topological. The metric is then identified with some components of the
gauge fields. Such constraints usually break the gauge group into a subgroup.
In the noncommutative field theoretic approach to gravity this works after the
constraints are imposed, provided that both the gauge group and the remaining
subgroup are of the unitary type. There is a formulation of noncommutative
gauge theories where the gauge group could also be of the orthogonal or
symplectic type, but it turned out that there are problems associated with
this formulation \cite{sheik},\cite{wess},\cite{deformed}. There is an
alternative interpretation in the case where the constraints could be solved
for some of the gauge fields in terms of the others. In this case one can
insist on preserving gauge invariance in a non-linear fashion, while changing
the gauge transformations of those gauge fields that are now dependent in such
a way as to preserve the constraints \cite{van}. In this paper we give an
invariant four-dimensional gravitational action and then generalize it to the
noncommutative case. The action is based on gauging the group $U(2,2)$ broken
by constraints to $U(1,1)\times U(1,1).$ One obtains, depending on the
constraints, topological gravity, Einstein gravity or conformal gravity. This
construction can be extended to the noncommutative case by replacing ordinary
products with star products. We derive the deformed curvatures, the deformed
action and compute corrections to first order in the deformation parameter
$\theta$ using the Seiberg-Witten map. We show that in this approach it is
only possible to deform Gauss-Bonnet topological gravity, or conformal gravity
but not Einstein gravity.

The noncommutative gravitational action was derived in dimensions two and
three \cite{6authors},\cite{nair},\cite{klemm}. In four-dimensions the
smallest unitary group that contains both the spin-connection and the vierbein
which spans the group $SO(1,4)$ or $SO(2,3)$ is $U(2,2)$ or $U(1,3).$ For
definiteness we will consider the group $U(2,2).$ The constraints should keep
the $SO(1,3)$ subgroup invariant. The appropriate subgroup is $U(1,1)\times
U(1,1).$ To be precise we define the $U(2,2)$ algebra as the set of $4\times4$
matrices $M$ satisfying \cite{physreport}
\[
g^{\dagger}\Gamma_{4}g=\Gamma_{4},
\]
where the $4\times4$ gamma matrices $\Gamma_{a},\quad a=1,2,3,4$ are the basis
of a Clifford algebra
\[
\left\{  \Gamma_{a},\Gamma_{b}\right\}  =2\delta_{ab},
\]
and where we have adopted the notation $\Gamma_{4}=i\Gamma_{0}$ and
$x^{4}=ix^{0}.$ The gauge fields $A_{\mu}$ satisfy
\[
A_{\mu}^{\dagger}=-\Gamma_{4}A_{\mu}\Gamma_{4}%
\]
and transform according to
\[
A_{\mu}^{g}=g^{-1}A_{\mu}g+g^{-1}\partial_{\mu}g.
\]
We can write
\[
A=\left(  ia_{\mu}+b_{\mu}\Gamma_{5}+e_{\mu}^{a}\Gamma_{a}+f_{\mu}^{a}%
\Gamma_{a}\Gamma_{5}+\frac{1}{4}\omega_{\mu}^{ab}\Gamma_{ab}\right)  dx^{\mu},
\]
where
\[
\Gamma_{5}=\Gamma_{1}\Gamma_{2}\Gamma_{3}\Gamma_{4},\quad\Gamma_{ab}=\frac
{1}{2}\left(  \Gamma_{a}\Gamma_{b}-\Gamma_{b}\Gamma_{a}\right)  .
\]
Let
\begin{align*}
D &  =d+A,\\
D^{2} &  =F=(dA+A^{2}),
\end{align*}
so that $F$ transforms covariantly $F^{g}=g^{-1}Fg$. Decomposing the field
strength in terms of the Clifford algebra generators
\[
F_{\mu\nu}=iF_{\mu\nu}^{1}+F_{\mu\nu}^{5}\Gamma_{5}+F_{\mu\nu}^{a}\Gamma
_{a}+F_{\mu\nu}^{a5}\Gamma_{a}\Gamma_{5}+\frac{1}{4}F_{\mu\nu}^{ab}\Gamma
_{ab},
\]
where $F=\frac{1}{2}F_{\mu\nu}dx^{\mu}\wedge dx^{\nu},$ then the components
are given by
\begin{align*}
F_{\mu\nu}^{1} &  =\partial_{\mu}a_{\nu}-\partial_{\nu}a_{\mu},\\
F_{\mu\nu}^{5} &  =\partial_{\mu}b_{\nu}-\partial_{\nu}b_{\mu}+2e_{\mu}%
^{a}f_{\nu a}-2e_{\nu}^{a}f_{\mu a},\\
F_{\mu\nu}^{a} &  =\partial_{\mu}e_{\nu}^{a}-\partial_{\nu}e_{\mu}^{a}%
+\omega_{\mu}^{ab}e_{\nu b}-\omega_{\nu}^{ab}e_{\mu b}+2f_{\mu}^{a}b_{\nu
}-2f_{\nu}^{a}b_{\mu},\\
F_{\mu\nu}^{a5} &  =\partial_{\mu}f_{\nu}^{a}-\partial_{\nu}f_{\mu}^{a}%
+\omega_{\mu}^{ab}f_{\nu b}-\omega_{\nu}^{ab}f_{\mu b}+2e_{\mu}^{a}b_{\nu
}-2e_{\nu}^{a}b_{\mu},\\
F_{\mu\nu}^{ab} &  =\partial_{\mu}\omega_{\nu}^{ab}+\omega_{\mu}^{ac}%
\omega_{\nu c}^{\quad b}+4\left(  e_{\mu}^{a}e_{\nu}^{b}-f_{\mu}^{a}f_{\nu
}^{b}\right)  -\mu\longleftrightarrow\nu,
\end{align*}
We can impose the constraints
\[
F_{\mu\nu}^{a}+F_{\mu\nu}^{a5}=0,\quad or\quad F_{\mu\nu}^{a}-F_{\mu\nu}%
^{a5}=0,
\]
which break the gauge group $U(2,2)$ to $U(1,1)\times U(1,1)$ with generators
\[
\left(  1\pm\Gamma_{5}\right)  \left\{  1,\,\Gamma_{ab}\right\}
\]
One can solve the above constraints to determine $\omega_{\mu}^{ab}$ in terms
of $e_{\mu}^{a\pm}=e_{\mu}^{a}\pm f_{\mu}^{a}$ and $b_{\mu}.$ We can rewrite
the constraints in the form
\[
\partial_{\mu}e_{\nu}^{a+}-\partial_{\nu}e_{\mu}^{a+}+\omega_{\mu\,b}%
^{a}e_{\nu}^{b+}-\omega_{\nu\,b}^{a}e_{\mu}^{b+}+2e_{\mu}^{a+}b_{\nu}-2e_{\nu
}^{a+}b_{\mu}=0,
\]
or
\[
\partial_{\mu}e_{\nu}^{a-}-\partial_{\nu}e_{\mu}^{a-}+\omega_{\mu\,b}%
^{a}e_{\nu}^{b-}-\omega_{\nu\,b}^{a}e_{\mu}^{b-}-2e_{\mu}^{a-}b_{\nu}+2e_{\nu
}^{a-}b_{\mu}=0,
\]
which imply that $\omega_{\mu}^{ab}=\omega_{\mu}^{ab}\left(  e_{\mu}%
^{a+},b_{\mu}\right)  $ or $\omega_{\mu}^{ab}=\omega_{\mu}^{ab}\left(  e_{\mu
}^{a-},-b_{\mu}\right)  .$ The solutions which recover the Einstein action are
obtained by imposing both sets of constraints simultaneously as these imply
\[
f_{\mu}^{a}=\alpha e_{\mu}^{a},\quad b_{\mu}=0,
\]
where $\alpha$ is an arbitrary parameter.

The action which is invariant under the remaining $U(1,1)\times U(1,1)$ group
is given by\ \cite{mcdowel},\cite{ortho},
\[
I=i\int\limits_{M}Tr\left(  \Gamma_{5}F\wedge F\right)
\]
where $F=\frac{1}{2}F_{\mu\nu}dx^{\mu}\wedge dx^{\nu}.$ Notice that
$\Gamma_{5}$ commutes with the generators $\left\{  1,\Gamma_{5},\Gamma
_{ab}\right\}  $ of $U(1,1)\times U(1,1)$ thus insuring the invariance of the
action. This action is metric independent, and one expects the space-time
metric to be generated from the gauge fields $e_{\mu}^{a}$ and $f_{\mu}^{a}.$
To see this we write the action when both sets of constraints are imposed
simultaneously and the only independent field is $e_{\mu}^{a}.$ The action
reduces to
\[
I=\frac{i}{4}\int\limits_{M}d^{4}x\epsilon^{\mu\nu\rho\sigma}\epsilon
_{abcd}\left(  R_{\mu\nu}^{ab}+8\left(  1-\alpha^{2}\right)  e_{\mu}^{a}%
e_{\nu}^{b}\right)  \left(  R_{\rho\sigma}^{cd}+8\left(  1-\alpha^{2}\right)
e_{\rho}^{c}e_{\sigma}^{d}\right)
\]
There are three possibilities $\left|  \alpha\right|  <1,$ $\left|
\alpha\right|  =1$ and $\left|  \alpha\right|  >1.$ \ The case $\left|
\alpha\right|  =1$ gives only the Gauss-Bonnet term and is topological. The
cases with $\left|  \alpha\right|  <1$ and $\left|  \alpha\right|  >1$ give
also the scalar curvature and cosmological constants with opposite signs. The
abelian gauge field $a_{\mu\text{ }}$decouples. This theory is different from
the usual gauge formulations in that it has more vacua, and it allows for
solutions with arbitrary cosmological constant. We could have restricted
ourselves to $SU(2,2)$ instead of $U(2,2)$ as the gauge field $a_{\mu}$
decouples, but we did not do so because such a choice is not allowed in the
noncommutative case. When only one of the constraints is imposed, then the
form of the action does not change, where $e_{\mu}^{a+}$ is taken to be the
independent field, we should solve for $e_{\mu}^{a-}$ from its equation of
motion. It is known that the action in this case gives conformal supergravity
\cite{physreport}.

We are now ready to deal with formulating an action for gravity which is
invariant under the star product. One of the main difficulties we mentioned in
previous work is that the metric defined by $g_{\mu\nu}=e_{\mu}^{a}\ast e_{\nu
a}$ is complex \cite{complex} and one has to obtain the correct action for the
non-symmetric part (or the complex part) of the metric \cite{ES}%
,\cite{schrod}. The other problem is related to finding an invariant measure
with respect to the star product \cite{NK}. Both of these problems could be
solved by adopting the formalism given above. We shall show that the deformed
vierbein $\widehat{e}_{\mu}^{a}$ remains real. Gauge invariance with
constraints eliminates some of the superfluous degrees of freedom. The
constraints also make it possible to have non-topological actions with the
advantage of not introducing a metric. The vierbeins are gauge fields
corresponding to the broken generators. The action being a $4$ form in $D=4$
dimensions is automatically invariant under the star product. The gauge fields
transform according to
\[
\widetilde{A}^{g}=\widetilde{g}_{\ast}^{-1}\ast\widetilde{A}\ast\widetilde
{g}+\widetilde{g}_{\ast}^{-1}\ast d\widetilde{g},
\]
where $\widetilde{g}$ satisfies
\[
\widetilde{g}_{\ast}^{-1}\ast\widetilde{g}=1,\quad\widetilde{g}^{\dagger}%
\ast\Gamma_{4}\ast\widetilde{g}=\Gamma_{4},
\]
and the gauge field strength is
\[
\widetilde{F}=(d\widetilde{A}+\widetilde{A}\ast\widetilde{A}),
\]
where
\[
\widetilde{A}=\widetilde{A}_{\mu}dx^{\mu},\quad\widetilde{F}=\frac{1}%
{2}\widetilde{F}_{\mu\nu}dx^{\mu}\wedge dx^{\nu},
\]
and the coordinates $x^{\mu}$ satisfy
\[
\left[  x^{\mu},x^{\nu}\right]  =i\theta^{\mu\nu},\quad\left[  \partial_{\mu
},\partial_{\nu}\right]  =0,\quad dx^{\mu}\wedge dx^{\nu}=-dx^{\nu}\wedge
dx^{\mu},
\]
which insures that $d^{2}=0.$ We use the property
\begin{align*}
\widetilde{A}\ast\widetilde{A}  &  =\widetilde{A}_{\mu}^{I}\ast\widetilde
{A}_{\nu}^{J}T_{I}T_{J}dx^{\mu}\wedge dx^{\nu}\\
&  =\frac{1}{2}\left(  \widetilde{A}_{\mu}^{I}\ast_{s}\widetilde{A}_{\nu}%
^{J}\left[  T_{I},T_{J}\right]  +\widetilde{A}_{\mu}^{I}\ast_{a}\widetilde
{A}_{\nu}^{J}\left\{  T_{I},T_{J}\right\}  \right)  dx^{\mu}\wedge dx^{\nu},
\end{align*}
where we have defined both the symmetric and antisymmetric star products by
\begin{align*}
f\ast_{s}g  &  \equiv\frac{1}{2}\left(  f\ast g+g\ast f\right)  =fg+\left(
\frac{i}{2}\right)  ^{2}\theta^{\mu\nu}\theta^{\kappa\lambda}\partial_{\mu
}\partial_{\kappa}f\partial_{\nu}\partial_{\lambda}g+O(\theta^{4}).\\
f\ast_{a}g  &  \equiv\frac{1}{2}\left(  f\ast g-g\ast f\right)  =\left(
\frac{i}{2}\right)  \theta^{\mu\nu}\partial_{\mu}f\partial_{\nu}g+\left(
\frac{i}{2}\right)  ^{3}\theta^{\mu\nu}\theta^{\kappa\lambda}\theta
^{\alpha\beta}\partial_{\mu}\partial_{\kappa}\partial_{\alpha}f\partial_{\nu
}\partial_{\lambda}\partial_{\beta}g+O(\theta^{5}).
\end{align*}
and $T_{I\text{ }}$are the Lie algebra generators. Notice that both
commutators and anticommutators appear in the products, making it necessary to
consider only the unitary groups. The advantage in using the Dirac matrix
representation is that all the generators corresponding to an even number of
gamma matrices form the subgroup $U(1,1)\times U(1,1)$ of $U(2,2)$ while the
generators corresponding to an odd number of gamma matrices belong to the
coset space $\frac{U(2,2)}{U(1,1)\times U(1,1)}$. Therefore one can constrain
some of the field strengths corresponding to the generators with an odd number
of gamma matrices to zero thus breaking the symmetry. It is \ more difficult
to solve the constraints in the noncommutative case. We shall make use of the
Seiberg-Witten map to do this. The S-W map is defined by the relation
\cite{SW}
\[
\widetilde{g_{\ast}}^{-1}\ast\widetilde{A}(A)\ast\widetilde{g}+\widetilde
{g_{\ast}}^{-1}\ast d\widetilde{g}=\widetilde{A}(g^{-1}Ag+g^{-1}dg),
\]
and whose solution is equivalent to \cite{SW}
\begin{align*}
\delta\widetilde{A}_{\mu}\left(  \theta\right)   &  =-\frac{i}{4}\delta
\theta^{\nu\rho}\left\{  \widetilde{A}_{\nu},\left(  \partial_{\rho}%
\widetilde{A}_{\mu}+\widetilde{F}_{\rho\mu}\right)  \right\}  _{\ast},\\
\delta\widetilde{\lambda}\left(  \theta\right)   &  =\frac{i}{4}\delta
\theta^{\nu\rho}\left\{  \partial_{\nu}\lambda,A_{\rho}\right\}  _{\ast},
\end{align*}
where we have defined $\widetilde{g}=e^{\widetilde{\lambda}}$ and
$g=e^{\lambda}.$ These transformations do not preserve the constraints. To
make these transformations compatible with the constraints one can follow the
same procedure as in the commutative case. This is done by first solving the
constraints and determining the dependent fields in terms of the independent
ones and then modifying the transformations of these dependent fields in such
a way as to preserve the constraints.

The constraints are given by
\[
\widetilde{F}_{\mu\nu}^{a}+\widetilde{F}_{\mu\nu}^{a5}=0,\quad or\quad
\widetilde{F}_{\mu\nu}^{a}-\widetilde{F}_{\mu\nu}^{a5}=0,
\]
and the action invariant under $U(1,1)\times U(1,1)$ is
\[
I=i\int\limits_{M}Tr\left(  \Gamma_{D+1}\widetilde{F}\ast\widetilde{F}\right)
.
\]
Notice that we can write $\widetilde{F}=\frac{1}{2}\widetilde{F}_{\mu\nu
}dx^{\mu}\wedge dx^{\nu}$ and $\widetilde{F}\ast\widetilde{F}=\frac{1}{2^{2}%
}\widetilde{F}_{\mu_{1}\mu_{2}}\ast\widetilde{F}_{\mu_{3}\mu_{4}}dx^{\mu_{1}%
}\wedge dx^{\mu_{2}}\wedge dx^{\mu_{3}}\wedge dx^{\mu_{4}}$ $.$ The gauge
fields $\widetilde{A}_{\mu}$ are decomposed as in the commutative case. The
field strengths are given by
\begin{align}
\widetilde{F}_{\mu\nu}\left(  1\right)   &  =i\left(  \partial_{\mu}%
\widetilde{a}_{\nu}-\partial_{\nu}\widetilde{a}_{\mu}\right) \nonumber\\
&  +2\left(  -\widetilde{a}_{\mu}\ast_{a}\widetilde{a}_{\nu}+\widetilde
{b}_{\mu}\ast_{a}\widetilde{b}_{\nu}+\widetilde{e}_{\mu}^{a}\ast_{a}%
\widetilde{e}_{\nu a}-\widetilde{f}_{\mu}^{a}\ast_{a}\widetilde{f}_{\nu
a}-\frac{1}{4}\widetilde{\omega}_{\mu}^{ab}\ast_{a}\widetilde{\omega
}_{\upsilon ab}\right)  ,\nonumber\\
\widetilde{F}_{\mu\nu}\left(  \Gamma_{5}\right)   &  =\partial_{\mu}%
\widetilde{b}_{\nu}-\partial_{\nu}\widetilde{b}_{\mu}+2\left(  \widetilde
{e}_{\mu}^{a}\ast_{s}\widetilde{f}_{\nu a}-\widetilde{f}_{\mu}^{a}\ast
_{s}\widetilde{e}_{\nu a}\right) \nonumber\\
&  +2\left(  \widetilde{b}_{\mu}\ast_{a}\widetilde{a}_{\nu}+\widetilde{a}%
_{\mu}\ast_{a}\widetilde{b}_{\nu}\right)  +\frac{1}{8}\epsilon_{abcd}%
\widetilde{\omega}_{\mu}^{ab}\ast_{a}\widetilde{\omega}_{\nu}^{cd},\nonumber\\
\widetilde{F}_{\mu\nu}\left(  \Gamma_{ab}\right)   &  =\frac{1}{4}\left(
\partial_{\mu}\widetilde{\omega}_{\nu}^{ab}-\partial_{\nu}\widetilde{\omega
}_{\mu}^{ab}+\widetilde{\omega}_{\mu}^{ac}\ast_{s}\widetilde{\omega}_{\nu
c}^{\quad b}-\widetilde{\omega}_{\mu}^{bc}\ast_{s}\widetilde{\omega}_{\nu
c}^{\quad a}\right) \nonumber\\
&  +\frac{i}{2}\left(  \widetilde{a}_{\mu}\ast_{a}\widetilde{\omega}_{\nu
}^{ab}+\widetilde{\omega}_{\mu}^{ab}\ast_{a}\widetilde{a}_{\nu}\right)
-\frac{1}{4}\epsilon_{\quad cd}^{ab}\left(  \widetilde{b}_{\mu}\ast
_{a}\widetilde{\omega}_{\nu}^{cd}+\widetilde{\omega}_{\mu}^{cd}\ast
_{a}\widetilde{b}_{\nu}\right) \nonumber\\
&  -4\epsilon_{\quad cd}^{ab}\left(  \widetilde{e}_{\mu}^{c}\ast_{a}%
\widetilde{f}_{\nu}^{d}+\widetilde{f}_{\mu}^{d}\ast_{a}\widetilde{e}_{\nu}%
^{c}\right)  +\left(  \widetilde{e}_{\mu}^{a}\ast_{s}\widetilde{e}_{\nu}%
^{b}-\widetilde{e}_{\nu}^{a}\ast_{s}\widetilde{e}_{\mu}^{b}-\widetilde{f}%
_{\mu}^{a}\ast_{s}\widetilde{f}_{\nu}^{b}+\widetilde{f}_{\nu}^{a}\ast
_{s}\widetilde{f}_{\mu}^{b}\right)  ,\nonumber
\end{align}
for the generators with an even number of gamma matrices, and by
\begin{align}
\widetilde{F}_{\mu\nu}\left(  \Gamma_{a}\right)   &  =\partial_{\mu}%
\widetilde{e}_{\nu}^{a}-\partial_{\nu}\widetilde{e}_{\mu}^{a}+\widetilde
{\omega}_{\mu}^{ac}\ast_{s}\widetilde{e}_{\nu c}+\widetilde{e}_{\mu}^{c}%
\ast_{s}\widetilde{\omega}_{\nu c}^{\quad a}\nonumber\\
&  -2\left(  \widetilde{b}_{\mu}\ast_{s}\widetilde{f}_{v}^{a}-\widetilde
{f}_{\mu}^{a}\ast_{s}\widetilde{b}_{\nu}\right)  +2i\left(  \widetilde{a}%
_{\mu}\ast_{a}\widetilde{e}_{\nu}^{a}+\widetilde{e}_{\mu}^{a}\ast
_{a}\widetilde{a}_{\nu}\right) \nonumber\\
&  +\frac{1}{2}\epsilon_{\,\,bcd}^{a}\left(  \widetilde{f}_{\mu}^{b}\ast
_{a}\widetilde{\omega}_{\nu}^{cd}+\widetilde{\omega}_{\mu}^{cd}\ast
_{a}\widetilde{f}_{\nu}^{b}\right)  ,\nonumber\\
\widetilde{F}_{\mu\nu}\left(  \Gamma_{a}\Gamma_{5}\right)   &  =\partial_{\mu
}\widetilde{f}_{\nu}^{a}-\partial_{\nu}\widetilde{f}_{\mu}^{a}+\widetilde
{\omega}_{\mu}^{ac}\ast_{s}\widetilde{f}_{\nu c}+\widetilde{f}_{\mu}^{c}%
\ast_{s}\widetilde{\omega}_{\nu c}^{\quad a}\nonumber\\
&  -2\left(  \widetilde{b}_{\mu}\ast_{s}\widetilde{e}_{v}^{a}-\widetilde
{e}_{\mu}^{a}\ast_{s}\widetilde{b}_{\nu}\right)  +2i\left(  \widetilde{a}%
_{\mu}\ast_{a}\widetilde{f}_{\nu}^{a}+\widetilde{f}_{\mu}^{a}\ast
_{a}\widetilde{a}_{\nu}\right) \nonumber\\
&  +\frac{1}{2}\epsilon_{\,\,\;bcd}^{a}\left(  \widetilde{e}_{\mu}^{b}\ast
_{a}\widetilde{\omega}_{\nu}^{cd}+\widetilde{\omega}_{\mu}^{cd}\ast
_{a}\widetilde{e}_{\nu}^{b}\right)  ,\nonumber
\end{align}
for the generators with an odd number of gamma matrices. In four dimensions,
the action is
\begin{align*}
I  &  =i\int\limits_{M}Tr\left(  \Gamma_{5}\widetilde{F}\ast\widetilde
{F}\right) \\
&  =i\int\limits_{M}d^{4}x\epsilon^{\mu\nu\rho\sigma}Tr\left(  \Gamma
_{5}\widetilde{F}_{\mu\nu}\ast\widetilde{F}_{\rho\sigma}\right) \\
&  =i\int\limits_{M}d^{4}x\epsilon^{\mu\nu\rho\sigma}\left(  2\widetilde
{F}_{\mu\nu}^{1}\ast_{s}\widetilde{F}_{\rho\sigma}^{5}+\epsilon_{abcd}%
\widetilde{F}_{\mu\nu}^{ab}\ast_{s}\widetilde{F}_{\rho\sigma}^{cd}\right)  .
\end{align*}
Notice that although only the symmetric star product appears there are linear
corrections in $\theta$ to the commutative action. As in the commutative case,
the constraints have to be solved for $\widetilde{\omega}_{\mu}^{ab}$ in terms
of $\widetilde{e}_{\mu}^{a+}$ or $\widetilde{e}_{\mu}^{a-}$ , $\widetilde
{b}_{\mu}$ and $\widetilde{a}_{\mu}.$ However, unlike the commutative case, it
is not possible to impose both constraints simultaneously after setting
$\widetilde{b}_{\mu}=0$ because of the presence of the $\pm e^{\pm}\omega$
term in $\widetilde{F}_{\mu\nu}^{a}\pm$ $\widetilde{F}_{\mu\nu}^{a5}$. These
two constraints become incompatible except in the special case where
$\widetilde{e}_{\mu}^{a-}=0,$ which corresponds to deforming the Gauss-Bonnet
action. If only one constraint is imposed and $\widetilde{\omega}_{\mu}^{ab}$
is determined from the constraint, the independent fields are $\widetilde
{e}_{\mu}^{a+}$ , $\widetilde{e}_{\mu}^{a-}$ , $\widetilde{b}_{\mu}$ and
$\widetilde{a}_{\mu}$ resulting in deformed conformal supergravity. It is not
possible to obtain a deformation of Einstein gravity as the constraints could
not be imposed simultaneously.\qquad

One can expand this action prerturbatively in powers of $\theta.$ This can be
done by using the Seiberg-Witten map for $\widetilde{e}_{\mu}^{a+}$
$\widetilde{e}_{\mu}^{a-}$ , $\widetilde{b}_{\mu}$ and $\widetilde{a}_{\mu}.$
These expressions are then used in the above constraint to determine
$\widetilde{\omega}_{\mu}^{ab}.$ It is instructive to carry this procedure to
first order in $\theta.$ Applying the Seiberg-Witten map, one gets
\begin{align*}
\widetilde{e}_{\mu}^{a\pm}  &  =e_{\mu}^{a\pm}+\frac{1}{2}\theta^{\kappa\rho
}\left(  a_{\kappa}\partial_{\rho}e_{\mu}^{a\pm}+e_{\kappa}^{a\pm}\left(
2\partial_{\rho}a_{\mu}-\partial_{\mu}a_{\rho}\right)  \right. \\
&  \qquad\qquad\left.  \mp\frac{i}{4}\epsilon_{abcd}\left(  e_{\kappa}^{b\pm
}\left(  \partial_{\rho}\omega_{\mu}^{cd}+F_{\rho\mu}^{cd}\right)
+\omega_{\kappa}^{cd}\partial_{\rho}e_{\mu}^{b\pm}\right)  \right)
+O(\theta^{2})\\
&  \equiv e_{\mu}^{a\pm}+\frac{1}{2}\theta^{\kappa\rho}e_{\mu\kappa\rho}%
^{a\pm}+O(\theta^{2})
\end{align*}%
\begin{align*}
\widetilde{a}_{\mu}  &  =a_{\mu}+\frac{1}{2}\theta^{\kappa\rho}\left(
a_{\kappa}\left(  2\partial_{\rho}a_{\mu}-\partial_{\mu}a_{\rho}\right)
-b_{\kappa}\left(  \partial_{\rho}b_{\mu}+F_{\rho\mu}^{5}\right)  \right. \\
&  \qquad\qquad\qquad-e_{\kappa}^{a}\left(  \partial_{\rho}e_{\mu}^{a}%
+F_{\rho\mu}^{a}\right)  +f_{\kappa}^{a}\left(  \partial_{\rho}f_{\mu}%
^{a}+F_{\rho\mu}^{a5}\right) \\
&  \qquad\qquad\qquad\left.  +\frac{1}{8}\omega_{\kappa}^{ab}\left(
\partial_{\rho}\omega_{\mu}^{ab}+F_{\rho\mu}^{ab}\right)  \right)
+O(\theta^{2})\\
&  \equiv a_{\mu}+\frac{1}{2}\theta^{\kappa\rho}a_{\mu\kappa\rho}+O(\theta
^{2})
\end{align*}%
\begin{align*}
\widetilde{b}_{\mu}  &  =b_{\mu}+\frac{1}{2}\theta^{\kappa\rho}\left(
b_{\kappa}\left(  2\partial_{\rho}a_{\mu}-\partial_{\mu}a_{\rho}\right)
+a_{\kappa}\left(  \partial_{\rho}b_{\mu}+F_{\rho\mu}^{5}\right)  \right. \\
&  \qquad\qquad\qquad\left.  -\frac{i}{8}\epsilon_{abcd}\omega_{\kappa}%
^{ab}\left(  \partial_{\rho}\omega_{\mu}^{cd}+F_{\rho\mu}^{cd}\right)
\right)  +O(\theta^{2})\\
&  \equiv b_{\mu}+\frac{1}{2}\theta^{\kappa\rho}b_{\mu\kappa\rho}+O(\theta
^{2})
\end{align*}

We do not take $\widetilde{\omega}_{\mu}^{ab}$ as given by the S-W map, but
instead substitute the above expressions in the constraint equation to
determine its value. First we write
\[
\widetilde{\omega}_{\mu}^{ab}=\omega_{\mu}^{ab}+\frac{1}{2}\theta^{\kappa\rho
}\omega_{\mu\kappa\rho}^{ab}+O(\theta^{2})
\]
then the constraint becomes
\begin{align*}
\widetilde{F}_{\mu\nu}^{a+}  &  =F_{\mu\nu}^{a+}+\frac{1}{2}\theta^{\kappa
\rho}\left(  \partial_{\mu}e_{\nu\kappa\rho}^{a+}-\partial_{\nu}e_{\mu
\kappa\rho}^{a+}+\omega_{\mu}^{ac}e_{\nu\kappa\rho}^{c+}-\omega_{\nu}%
^{ac}e_{\mu\kappa\rho}^{c+}\right. \\
&  \qquad\qquad\qquad+\omega_{\mu\kappa\rho}^{ac}e_{\nu}^{c+}-\omega
_{\nu\kappa\rho}^{ac}e_{\mu}^{c+}\mp2\left(  b_{\mu\kappa\rho}e_{\nu}%
^{a+}-b_{\nu\kappa\rho}e_{\mu}^{a+}\right) \\
&  \qquad\qquad\qquad\left.  -2\left(  \partial_{\kappa}a_{\mu}\partial_{\rho
}e_{\nu}^{a+}-\partial_{\kappa}a_{\nu}\partial_{\rho}e_{\mu}^{a+}\right)
\right)  +O(\theta^{2})
\end{align*}
Substituting $\widetilde{F}_{\mu\nu}^{a+}=0,$ and $F_{\mu\nu}^{a+}=0$, we can
solve for $\omega_{\mu\kappa\rho}^{ab}$ to obtain:
\[
\omega_{\mu\kappa\rho}^{ab}=\frac{1}{2}\left(  e^{\nu b+}C_{\mu\nu\kappa\rho
}^{a}-e^{\nu a+}C_{\mu\nu\kappa\rho}^{b}+e^{\sigma a+}e^{\nu b+}e_{\mu c}%
^{+}C_{\sigma\nu\kappa\rho}\right)
\]
where
\begin{align*}
C_{\mu\nu\kappa\rho}^{a}  &  =-\left(  \partial_{\mu}e_{\nu\kappa\rho}%
^{a+}-\partial_{\nu}e_{\mu\kappa\rho}^{a+}+\omega_{\mu}^{ac}e_{\nu\kappa\rho
}^{c+}-\omega_{\nu}^{ac}e_{\mu\kappa\rho}^{c+}\right. \\
&  \qquad\left.  -2\left(  \partial_{\kappa}a_{\mu}\partial_{\rho}e_{\nu}%
^{a+}-\partial_{\kappa}a_{\nu}\partial_{\rho}e_{\mu}^{a+}\right)  \right)
\end{align*}
To find the deformed action we first calculate
\begin{align*}
\widetilde{F}_{\mu\nu}^{1}  &  =F_{\mu\nu}^{1}+\frac{1}{2}\theta^{\kappa\rho
}\left(  \partial_{\mu}a_{\nu\kappa\rho}-\partial_{\nu}a_{\mu\kappa\rho
}-\partial_{\kappa}a_{\mu}\partial_{\rho}a_{\nu}+\partial_{\kappa}b_{\mu
}\partial_{\rho}b_{\nu}\right. \\
&  \qquad\qquad\qquad\left.  +\frac{1}{2}\left(  \partial_{\kappa}e_{\mu}%
^{a+}\partial_{\rho}e_{\nu}^{a-}-\partial_{\kappa}e_{\nu}^{a+}\partial_{\rho
}e_{\mu}^{a-}\right)  -\frac{1}{4}\partial_{\kappa}\omega_{\mu}^{ab}%
\partial_{\rho}\omega_{\nu}^{ab}\right)  +O(\theta^{2})\\
&  \equiv F_{\mu\nu}^{1}+\frac{1}{2}\theta^{\kappa\rho}F_{\mu\nu\kappa\rho
}^{1}+O(\theta^{2})
\end{align*}%
\begin{align*}
\widetilde{F}_{\mu\nu}^{ab}  &  =F_{\mu\nu}^{ab}+\frac{1}{2}\theta^{\kappa
\rho}\left(  \partial_{\mu}\omega_{\nu\kappa\rho}^{ab}-\partial_{\mu}%
\omega_{\nu\kappa\rho}^{ab}+\omega_{\mu}^{ac}\omega_{\nu\kappa\rho}%
^{cb}-\omega_{\nu}^{ac}\omega_{\mu\kappa\rho}^{cb}-\omega_{\mu}^{bc}%
\omega_{\nu\kappa\rho}^{ca}+\omega_{\nu}^{bc}\omega_{\mu\kappa\rho}%
^{ca}\right. \\
&  \qquad\qquad\qquad+4\left(  e_{\mu}^{a+}e_{\nu\kappa\rho}^{a-}-e_{\nu}%
^{a+}e_{\mu\kappa\rho}^{a-}-e_{\mu}^{a-}e_{\nu\kappa\rho}^{a+}+e_{\nu}%
^{a-}e_{\mu\kappa\rho}^{a+}\right) \\
&  \qquad\qquad\qquad-8i\epsilon_{abcd}\left(  \partial_{\kappa}e_{\mu}%
^{c+}\partial_{\rho}e_{\nu}^{d-}-\partial_{\kappa}e_{\nu}^{c+}\partial_{\rho
}e_{\mu}^{d-}\right)  -2\left(  \partial_{\kappa}a_{\mu}\partial_{\rho}%
\omega_{\nu}^{ab}-\partial_{\kappa}a_{\nu}\partial_{\rho}\omega_{\mu}%
^{ab}\right) \\
&  \qquad\qquad\qquad\left.  -i\epsilon_{abcd}\left(  \partial_{\kappa}b_{\mu
}\partial_{\rho}\omega_{\nu}^{cd}-\partial_{\kappa}b_{\nu}\partial_{\rho
}\omega_{\mu}^{cd}\right)  \right)  +O(\theta^{2})\\
&  \equiv F_{\mu\nu}^{ab}+\frac{1}{2}\theta^{\kappa\rho}F_{\mu\nu\kappa\rho
}^{ab}+O(\theta^{2})
\end{align*}
Notice that all the above expressions are real. The appearance of
$i\epsilon_{abcd}$ is due to the convention $x^{4}=ix^{0}$ so that
$i\epsilon_{1234}=\epsilon_{1230}=1.$ Therefore the conformal gravity action
to first order in $\theta$ is given by
\[
I=i\int d^{4}x\epsilon^{\mu\nu\lambda\sigma}\left(  \epsilon_{abcd}F_{\mu\nu
}^{ab}F_{\lambda\sigma}^{cd}+\theta^{\kappa\rho}\left(  2e_{\mu}^{a+}e_{\nu
}^{a-}F_{\lambda\sigma\kappa\rho}^{1}+\epsilon_{abcd}F_{\mu\nu}^{ab}%
F_{\lambda\sigma\kappa\rho}^{cd}\right)  \right)  +O(\theta^{2})
\]
where we have dropped total derivative terms. The deformation to the
Gauss-Bonnet action is obtained from the above expression by setting $e_{\mu
}^{a-}=0.$ It would be instructive to compare this action with the one
obtained from the Born-Infeld effective action in String theory where the
field $B_{\mu\nu}$ has a constant background \cite{SL}. One can also compare
these results by following the results of Jackiw-Pi \cite{JP} by defining
covariant coordinate transformations on noncommutative spaces. More
importantly is to compare this result with the spectral action for a deformed
spectral triple $(\widetilde{\mathcal{A}},\widetilde{H},\widetilde{D})$ where
$\widetilde{\mathcal{A}}=l(\mathcal{A}),$ $l$ is the left twist operator
\cite{CL}. The difficult part is to obtain the deformed operator
$\widetilde{D}$ and it is hoped that the above formulation will give some
hints on how to find the appropriate Dirac operator.

\ To summarize, we have proposed a four-dimensional gravitational action valid
for both commutative and noncommutative field theories. This action differs
from the familiar gravitational action in that it allows for other vacua
besides those of the metric theory. The noncommutativity is obtained by
replacing ordinary products with star products. The action is gauge invariant
and do not involve explicit use of the metric. Only conformal gravity or
Gauss-Bonnet topological gravity could be generalized to the noncommutative
case as the constraints imposed on the gauge field strengths should be
self-consistent. For some of the vacuum solutions, one of the gauge fields is
identified with the vierbein, and the theory becomes metric. It will be
interesting to study how to generalize this proposal to higher dimensions.
There are no fundamental obstacles to this approach in even dimensions. In odd
dimensions, however, it is not possible to impose constraints in such a way as
to preserve a smaller unitary group including the spin-connection generators
of $SO(2n+1).$ It appears that in odd dimensions the only gravitational
actions which are generalizable to the noncommutative case are of the
Chern-Simons type \cite{CF},\cite{poly}, and therefore must be topological.
Finally, one can study the supersymmetric version of the four-dimensional
gravitational action by considering the graded Lie-algebra $U(2,2|1).$

\section{Acknowledgments}

I\ would like to thank the Alexander von Humboldt Foundation for support
through a research award.

\end{document}